\documentclass[aps,prx,twocolumn,superscriptaddress,showpacs,longbibliography]{revtex4-1}
\usepackage{amsmath}
\usepackage{graphicx}
\usepackage{amsfonts}
\usepackage{verbatim}
\usepackage{amssymb}
\usepackage{color}
\usepackage{amsmath} % or simply amstext
\usepackage{hyperref}
\hypersetup{colorlinks = true, urlcolor = blue, linkcolor = blue, citecolor = blue}

\newcommand{\angstrom}{\text{\normalfont\AA}}

\newcommand{\su}{\uparrow} 
\newcommand{\sd}{\downarrow} 
\newcommand{\bpm}{\begin{pmatrix}}
\newcommand{\epm}{\end{pmatrix}}

\newcommand{\nn}{\nonumber \\} 
\newcommand{\tp}{ ^{\intercal} }

\begin{document}

\title{Electronic properties of InAs/EuS/Al hybrid nanowires}
%\title{Topological phases in InAs/EuS/Al hybrid nanowires}

\author{Chun-Xiao Liu}
\email{Electronic address: chunxiaoliu62@gmail.com}
\affiliation{Qutech, Delft University of Technology, Delft 2600 GA, The Netherlands.}
\affiliation{Kavli Institute of Nanoscience, Delft University of Technology, Delft 2600 GA, The Netherlands.}

\author{Sergej Schuwalow}
\affiliation{Center for Quantum Devices, Niels Bohr Institute, University of Copenhagen and Microsoft Quantum Materials Lab Copenhagen, Lyngby, Denmark.}

\author{Yu Liu}
\affiliation{Center for Quantum Devices, Niels Bohr Institute, University of Copenhagen and Microsoft Quantum Materials Lab Copenhagen, Lyngby, Denmark.}

\author{Kostas Vilkelis}
\affiliation{Qutech, Delft University of Technology, Delft 2600 GA, The Netherlands.}
\affiliation{Kavli Institute of Nanoscience, Delft University of Technology, Delft 2600 GA, The Netherlands.}

\author{A. L. R. Manesco}
\affiliation{Computational Materials Science Group (ComputEEL), Escola de Engenharia de Lorena, Universidade de S$\tilde{a}$o Paulo (EEL-USP), Materials Engineering Department (Demar), Lorena -- SP, Brazil}
\affiliation{Kavli Institute of Nanoscience, Delft University of Technology, Delft 2600 GA, The Netherlands.}

\author{P. Krogstrup}
\affiliation{Center for Quantum Devices, Niels Bohr Institute, University of Copenhagen and Microsoft Quantum Materials Lab Copenhagen, Lyngby, Denmark.}

\author{Michael Wimmer}
\affiliation{Qutech, Delft University of Technology, Delft 2600 GA, The Netherlands.}
\affiliation{Kavli Institute of Nanoscience, Delft University of Technology, Delft 2600 GA, The Netherlands.}

\date{\today}

\begin{abstract}
We study the electronic properties of InAs/EuS/Al heterostructures as explored in a recent experiment [S. Vaitiek\.enas \emph{et al.}, \href{https://doi.org/10.1038/s41567-020-1017-3}{Nat. Phys. (2020)}], combining both spectroscopic results and microscopic device simulations.  
In particular, we use angle-resolved photoemission spectroscopy to investigate the band bending at the InAs/EuS interface.
The resulting band offset value serves as an essential input to subsequent microscopic device simulations, allowing us to map the electronic wave function distribution.
We conclude that the magnetic proximity effects at the Al/EuS as well as the InAs/EuS interfaces are both essential to achieve topological superconductivity at zero applied magnetic field.
Mapping the topological phase diagram as a function of gate voltages and proximity-induced exchange couplings, we show that the ferromagnetic hybrid nanowire with overlapping Al and EuS layers can become a topological superconductor within realistic parameter regimes, and that the topological phase can be optimized by external gating.
Our work highlights the need for a combined experimental and theoretical effort for faithful device simulations.
\end{abstract}

\maketitle

\section{Introduction}

Topological superconductivity (TSC) has attracted lots of attention and inspired intensive research over the last few decades.
The defects or wire ends of a TSC can host Majorana zero modes which are non-Abelian anyons and potential building blocks of topological quantum computing~\cite{Nayak2008Non-Abelian, Alicea2012New, Leijnse2012Introduction, Beenakker2013Search, Stanescu2013Majorana, Jiang2013Non, Elliott2015Colloquium, DasSarma2015Majorana, Sato2016Majorana, Sato2017Topological, Aguado2017Majorana, Lutchyn2018Majorana, Zhang2019Next, Frolov2020Topological}.
Heterostructures between a spin-orbit coupled semiconducting nanowire and a conventional $s$-wave superconductor is one of the promising platforms for realizing TSC~\cite{Sau2010Generic, Lutchyn2010Majorana, Oreg2010Helical, Sau2010NonAbelian}.
In these hybrid devices, topological superconductivity is realized for a sufficiently strong Zeeman splitting.

\begin{figure}
\centering
{\includegraphics[width = \linewidth]{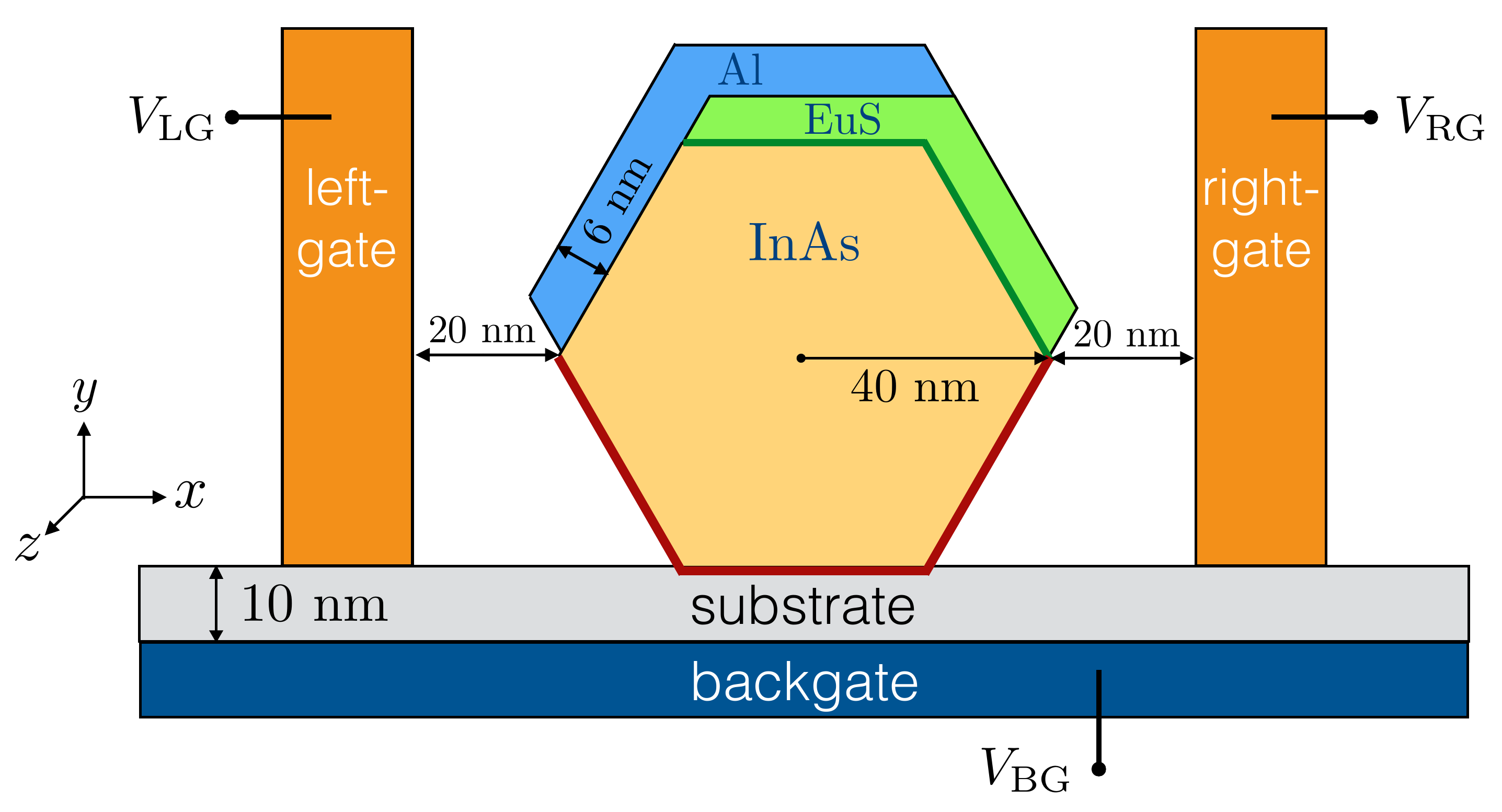}}
\caption{Schematic of the device studied in the experiment~\cite{Vaitiekenas2020Zerobias} and in this work. An InAs nanowire (yellow) is partially covered by Al (blue) and EuS (green) layers and is placed on a dielectric substrate (grey). A back-gate (dark blue) and two side-gates (orange) are applied to control the electrostatic potential profile in the InAs nanowire. Surface charges are added on the three facets of the bare InAs nanowire (brown) and on the two facets of the InAs/EuS interface (dark green) to account for the band bending effect.}
\label{fig:schematic}
\end{figure}

In most experimental studies of semiconductor-superconductor hybrid nanowires so far, Zeeman splitting is induced by an externally applied magnetic field~\cite{Mourik2012Signatures, Das2012Zero, Deng2012Anomalous,Churchill2013Superconductor, Finck2013Anomalous, Albrecht2016Exponential, Chen2017Experimental, Deng2016Majorana, Zhang2017Ballistic, Gul2018Ballistic, Nichele2017Scaling}. 
However, Zeeman energy in the hybrid system can also be induced by proximity effect from ferromagnetic insulators (FMI)~\cite{Sau2010Generic, Sau2010NonAbelian}.
In a recent experiment, topological properties of InAs/EuS/Al ferromagnetic (FM) hybrid nanowires have been investigated~\cite{Vaitiekenas2020Zerobias}. 
Tunneling spectroscopy revealed zero-bias conductance peaks over a finite parameter regime for multiple devices, compatible with Majorana zero modes and topological superconductivity.
Interestingly, such zero-bias peaks have appeared only in devices of a particular geometry, namely when the Al and EuS layers overlap with each other by one facet (see Fig.~\ref{fig:schematic}), but not in other device geometries without such an overlap.
This raises the question on the fundamental physical mechanisms for realizing TSC in such ferromagnetic hybrid nanowires.

In this work, we explore systematically different mechanisms for inducing an effective Zeeman energy in the nanowire, using detailed microscopic device simulations.
To this end it is essential to have a faithful description of the electrostatic potential in the device.
Previous works highlighted the critical role of band offsets at interfaces of the semiconductor with other materials~\cite{Antipov2018Effects, Mikkelsen2018Hybridization}.
For the bare InAs surface and the InAs/Al interface this has been studied systematically using angle-resolved photoemission spectroscopy (ARPES) \cite{Schuwalow2019Band}, but no such analysis has been available for the InAs/EuS interface so far.

We combine an analysis of the band offset at the InAs/EuS interface from ARPES data with electrostatic device simulations to arrive at a faithful description of the electronic density in these hybrid nanowires.
In particular, we find that the enhanced band bending at the InAs/EuS interface leads to an accumulation of electrons along these facets.
Using a microscopic model for superconductivity we conclude that the magnetic proximity effects at the Al/EuS as well as the InAs/EuS interfaces are both essential for inducing a sufficiently large effective Zeeman spin splitting allowing to reach a topological phase.
Our calculations show that a topological phase can be reached with plausible parameter values, and we discuss how topological properties can be optimized by external gating.

\section{Band bending and Electrostatics}

\subsection{Band bending at the InAs/EuS interface}
\begin{figure}
\centering
{\includegraphics[width = \linewidth]{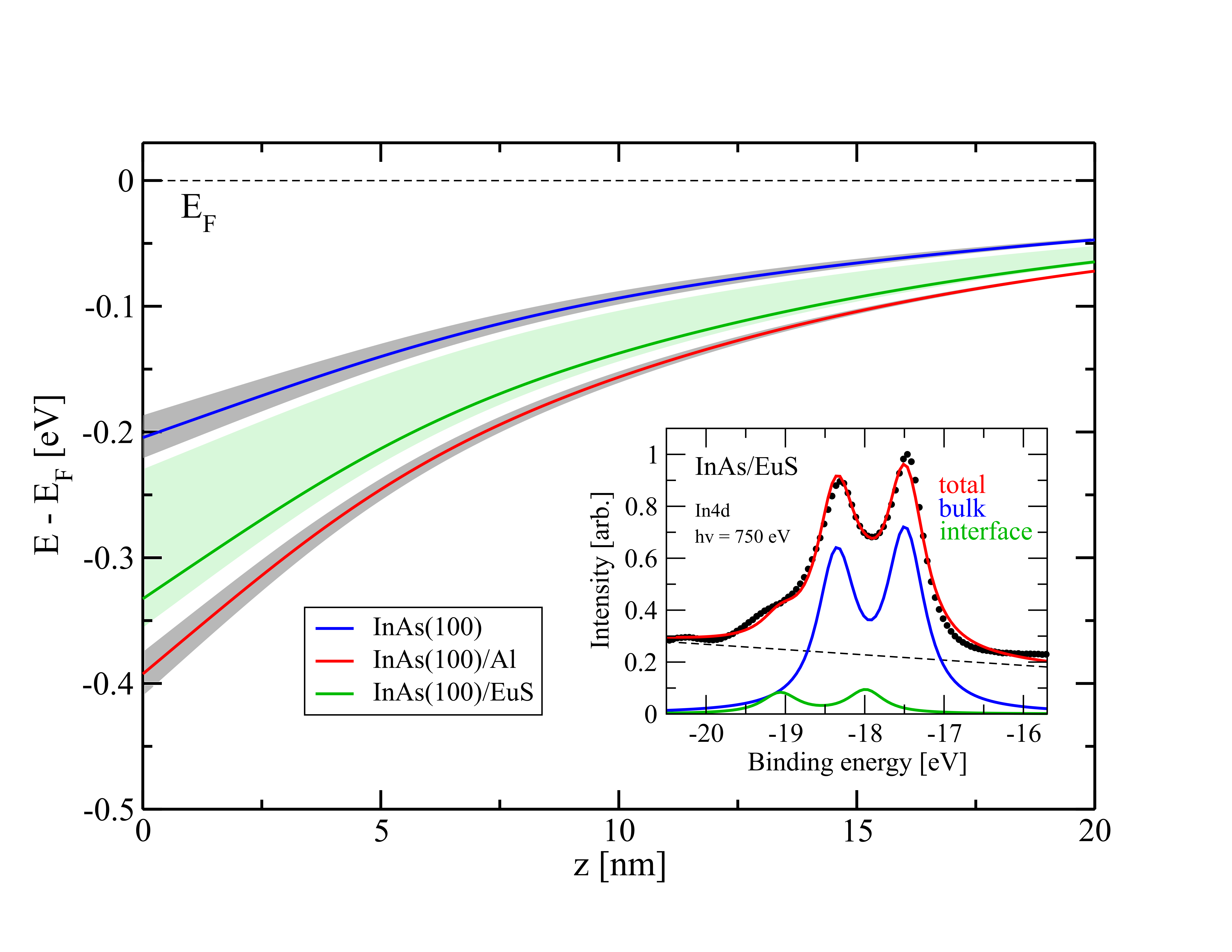}}
\caption{Interface band offsets and band bending profiles for the bare InAs(100) planar surface, the InAs(100)/Al, and InAs(100)/EuS heterostructures. 
Estimated confidence intervals are shown in grey and light green, respectively. 
Inset: Fit of the In4d core-level peaks of the InAs/EuS heterostructure for photon energy $h \nu$ = 750~eV.
The InAs(100)/EuS interface was grown in the MBE system of the Niels Bohr Institute in Copenhagen and transported for spectroscopic measurements at the ADRESS beamline of the SWISS Light Source at PSI, Switzerland in protective atmosphere.
Data for InAs and InAs/Al is from Ref.~\cite{Schuwalow2019Band}, and ARPES data obtained for InAs/EuS interface is in Ref.~\cite{Liu2020Coherent}.
}
\label{fig:band_offset}
\end{figure}

Accurate values of band offset at the interface of InAs with other materials are crucial for obtaining faithful electrostatic potential and charge density profiles inside the InAs nanowire.
In a previous work~\cite{Schuwalow2019Band}, the planar interfaces of InAs/Al and InAs/vacuum were both carefully investigated using the ARPES measurements along with the core-level fitting procedure.
The resulting values of the band offset of InAs(100)/Al and InAs(100)/vacuum, and the band bending profile near the interface are summarized as the blue and red lines in Fig.~\ref{fig:band_offset} (data from Ref.~\cite{Schuwalow2019Band}).

In this work, we focus on the band bending effect at the InAs(100)/EuS interface.
ARPES data obtained for this interface has been presented in Ref.~\cite{Liu2020Coherent}.
Here, we use the methods described in Ref.~\cite{Schuwalow2019Band} to extract the band bending from this data.
In particular, the fit of the In4d core-level spectra for the InAs/EuS interface is performed simultaneously for a set of photon energies in the range 350-750~eV.
We use a bulk and an interface component consisting of two Voigt functions each.
The broadening and shift of the line profile by the band bending potential is accounted for using an auxiliary Schr\"{o}dinger-Poisson simulation and the characteristic energy between the conduction band minimum and the In4d core level $\rm{\Delta}_{\rm{CL}}(\rm{In4d}, \rm{InAs}) = -17.22(3)$ eV for InAs~\cite{Schuwalow2019Band}.

The result of the core-level fitting for $h \nu$ = 750~eV is shown in the inset of Fig.~\ref{fig:band_offset}. 
While the overall shape of the core line is well captured by our model, the bulk component binding energy seems to be underestimated by $\sim$0.08~eV. 
We suspect that this may be caused by nonlinear behavior of the background or by a small additional interface component that is not adequately captured in our approach, which is reflected in the increased estimate for the confidence interval towards lower binding energies.
The bend bending profile of InAs(100)/EuS interface is shown as the green line in Fig.~\ref{fig:band_offset}, and we see that the band offset value of InAs/EuS is in between the values of InAs/Al and InAs/vacuum.

Finally, we note that owing to the geometrical difference between a planar interface and a multi-facet
nanowire structure, the band offset values shown in Fig.~\ref{fig:band_offset} should be regarded as guiding values. 
For the InAs/Al interface specifically, we typically observe the value of band offset for in-situ planar MBE growth shown here to be an upper bound, with a reduction of 0.05-0.1~eV for interfaces with a reduced quality using other growth modes such as growth after decapping. 
We can expect this to apply to growth on nanowire facets.
So without loss of generality, in this work we choose the band offset values in our model to be $W_{\rm{InAs/vac}}=$0.2~eV, $W_{\rm{InAs/EuS}}=$0.26~eV and $W_{\rm{InAs/Al}}=$0.35~eV, respectively.

\subsection{Thomas Fermi-Poisson approach}
The setup for studying the electrostatics in this work is schematically shown in Fig.~\ref{fig:schematic}. We focus on the two-dimensional cross section (in the $x$-$y$ plane) of the system, and assume translational symmetry along the third dimension ($z$ axis).
The hexagonal InAs nanowire of radius 40~nm is covered by the EuS layer on two of the top facets, and also covered by the Al layer on one adjacent facet. 
The hybrid nanowire is placed on a dielectric layer of thickness 10~nm, and a back-gate and two side-gates are applied below or beside the nanowire.
To obtain the electrostatic potential $\phi(\bold{r})$ for the setup, we solve the self-consistent Thomas Fermi-Poisson equation~\cite{Vuik2016Effects, Antipov2018Effects, Mikkelsen2018Hybridization, Woods2018Effective, Winkler2019Unified, Armagnat2019self-consistent}
\begin{align}
\nabla \cdot [ \varepsilon_r(\bold{r}) \nabla \phi(\bold{r}) ] =  \frac{\rho_{\rm{tot}}[ \phi(\bold{r}) ]}{\varepsilon_0},
\label{eq:TFP}
\end{align}
with appropriate boundary conditions.
Here the total charge density 
\begin{align}
\rho_{\rm{tot}}[\phi(\bold{r})]= \rho_{\text{e}}(\phi) + \rho_{\text{hh}}(\phi) + \rho_{\text{lh}}(\phi) + \rho_{\rm{surf}}
\end{align}
includes the conduction electrons, the heavy/light holes, and the surface charges. 
We use the Thomas-Fermi approximation for a 3D electron gas to determine the mobile charge densities inside the InAs nanowire: 
\begin{align}
&\rho_{\text{e}}(\phi) =  -\frac{e}{3\pi^2 } \left( \frac{2m_{\text{e}} e \phi \theta(\phi)}{\hbar^2}  \right)^{3/2}, \nn
&\rho_{\text{hh}/\text{lh}}(\phi) =  \frac{e}{3\pi^2} \left( \frac{ 2m_{\text{hh}/\text{lh}} (-e\phi-E_g) \theta(-e\phi-E_g)}{\hbar^2}  \right)^{3/2}
\end{align}
where $m_{\rm{e}}=0.023~m_{\text{0}}$, $m_{\text{hh}}=0.41~m_{\text{0}}$, $m_{\text{lh}}=0.026~m_{\text{0}}$ are the effective mass of the conduction electron, the heavy-hole and the light-hole in unit of electron mass, $E_g=0.418~$eV is the band gap between conduction and valence bands~\cite{Winkler2003Spin}, and $\theta(x)$ is the Heaviside step function. 
The surface charges are added to account for the band bending effect at both InAs/EuS and InAs/vacuum interfaces.
At the two top facets of the InAs nanowire, where it is in contact with the EuS layer, a positive charge layer of 1~nm thickness and density $\rho_{\rm{surf}}=1.8 \times 10^{19}~e/\rm{cm}^3$ is added, leading to a band offset $W_{\rm{InAs/EuS}} = 0.26~$eV.
Similarly, at the three facets where the InAs nanowire is either in contact with vacuum or the dielectric layer, another 1~nm thick positive charge layer of density $\rho_{\rm{surf}}=1.3 \times 10^{19}~e/\rm{cm}^3$ is applied to model the band offset value $W_{\rm{InAs/vac}} = 0.2~$eV~\cite{Olsson1996Charge, Degtyarev2017Features, Winkler2019Unified, Woods2020Subband}.
The band bending effect at the interface of InAs and the metallic aluminum layer is modeled by the Dirichlet boundary condition, i.e., $\phi = e^{-1} W_{\rm{InAs/Al}} = 0.35~$V at the remaining one facet of the InAs nanowire.
Additionally, the regions of the gates are also Dirichlet boundary conditions, with the values being determined by the applied voltage value, i.e., $\phi = V_{i}$, $i=$BG, LG, and RG.
It is noteworthy that the treatment of the band bending effect at the InAs/EuS interface is unique to this work, and thus distinguishes our work from others~\cite{Woods2020Electrostatic}

\section{Electronic properties of ferromagnetic hybrid nanowires}

\begin{table}[t]
\centering
\begin{tabular}{ p{3cm} p{2.5cm} p{2.5cm} }
\multicolumn{3}{ c }{Table I. Physical parameters for InAs and Al} \\
\hline \hline
Parameter (unit) & InAs & Al \\
\hline
m ($m_0$)  & 0.023~\cite{Winkler2003Spin} & 1 \\
$\alpha_R$ (eV\angstrom)  & 0.3~\cite{Gmitra2016First-principles}  & 0 \\
$E_F$(eV) & 0 & 11.27~\cite{Cochran1958Superconducting} \\ 
$\Delta_0$ (meV) & 0 & 0.34~\cite{Cochran1958Superconducting} \\
$\varepsilon_r$  & 15.15 &  \\
\hline \hline
\end{tabular}
\end{table}

\subsection{Model Hamiltonian}
The quantum model for investigating the electronic properties of the hybrid nanowire is shown in Fig.~\ref{fig:schematic}.
We consider the two-dimensional cross section of the nanowire ($xy$-plane), assuming translational symmetry along the wire axis ($z$-axis). 
The quantum system consists of only the InAs nanowire and the Al layer, which we treat on equal footing at the quantum mechanical level.
We model the role of EuS as an induced exchange coupling term in InAs and Al, while neglecting the stray field from EuS~\cite{Liu2020Semiconductor}.
The effects of gates, surface charges, dielectric layers, and the vacuum are taken into account via the self-consistently calculated electrostatic potential inside the InAs nanowire.
Under these assumptions, the normal-state Hamiltonian for the ferromagnetic hybrid nanowire can be written as
\begin{align}
H_{\rm{N}} = & \bold{p}\tp \frac{1}{2m(\bold{r})} \bold{p} + \alpha_R(\bold{r}) ( -i \partial_x \sigma_z - k_z \sigma_x ) - E_F(\bold{r}) \nn
& - e\phi(\bold{r}) + h_{\text{ex}}(\bold{r}) \sigma_z,
\label{eq:normal_ham}
\end{align}
where $\bold{p} = (-i \hbar \partial_x, -i\hbar \partial_y, \hbar k_z)$ is the momentum operator with $\hbar$ being the Planck constant, $k_z$ the wave vector along the nanowire axis, $\sigma_{i}$ the Pauli matrices acting on the spin space, $m(\bold{r})$ the effective mass, $\alpha_R(\bold{r})$ the strength of the Rashba spin-orbit coupling, $E_F(\bold{r})$ the Fermi energy, $\phi(\bold{r})$ the electrostatic potential, $e>0$ the elementary charge, and $h_{\text{ex}}(\bold{r})$ the strength of the induced exchange coupling due to the magnetic proximity effect from EuS. 
The physical parameters for InAs and Al are summarized in Table I. 
In addition, a random onsite potential is added within a distance of $2~$nm from the outer surface of Al, modeling the effect of disorder induced by the amorphous oxide layer in realistic devices~\cite{Antipov2018Effects}. 
We assume that the disorder potential has strength $U_0 = 1~$eV with zero average, and is spatially uncorrelated, i.e., $\langle \langle \delta E_F(\bold{r}) \rangle \rangle=0$, $\langle \langle \delta E_F(\bold{r}_i)\delta E_F(\bold{r}_j) \rangle \rangle = U^2_0 /3 \cdot \delta_{ij} $, such that the bands in Al and InAs couple to each other strongly~\cite{ Antipov2018Effects, Winkler2019Unified}.

When superconductivity is taken into consideration, the system is described by the Bogoliubov-de Gennes (BdG) Hamiltonian
\begin{align}
H_{\text{BdG}} = & \Big( \bold{p}\tp \frac{1}{2m(\bold{r})} \bold{p} + \alpha_R(\bold{r}) ( -i \partial_x \sigma_z - k_z \sigma_x ) - E_F(\bold{r}) \nn
&  - e\phi(\bold{r}) \Big) \tau_z + h_{\text{ex}}(\bold{r}) \sigma_z + \Delta(\bold{r}) \tau_x,
\label{eq:BdG_ham}
\end{align}
in the basis of $(\psi_{\rm{e} \su}, \psi_{\rm{e} \sd}, \psi_{\rm{h} \sd}, -\psi_{\rm{h} \su})$. Here $\tau_{i}$ are the Pauli matrices acting on the Nambu space, and $\Delta(\bold{r})$ is the pairing potential in the superconductor.

For the numerical calculations, the Hamiltonians in Eqs.~\eqref{eq:normal_ham} and~\eqref{eq:BdG_ham} are first discretized into a tight-binding model on a square lattice using the KWANT package~\cite{kwant}.
We choose the lattice constants for InAs and Al to be 5~$\angstrom$ and 1~$\angstrom$, respectively, to account for the large Fermi energy difference between the two materials.
Then the eigenenergies and eigenstates are obtained by diagonalizing the sparse Hamiltonian matrices.

\begin{figure}
\centering
{\includegraphics[width = \linewidth]{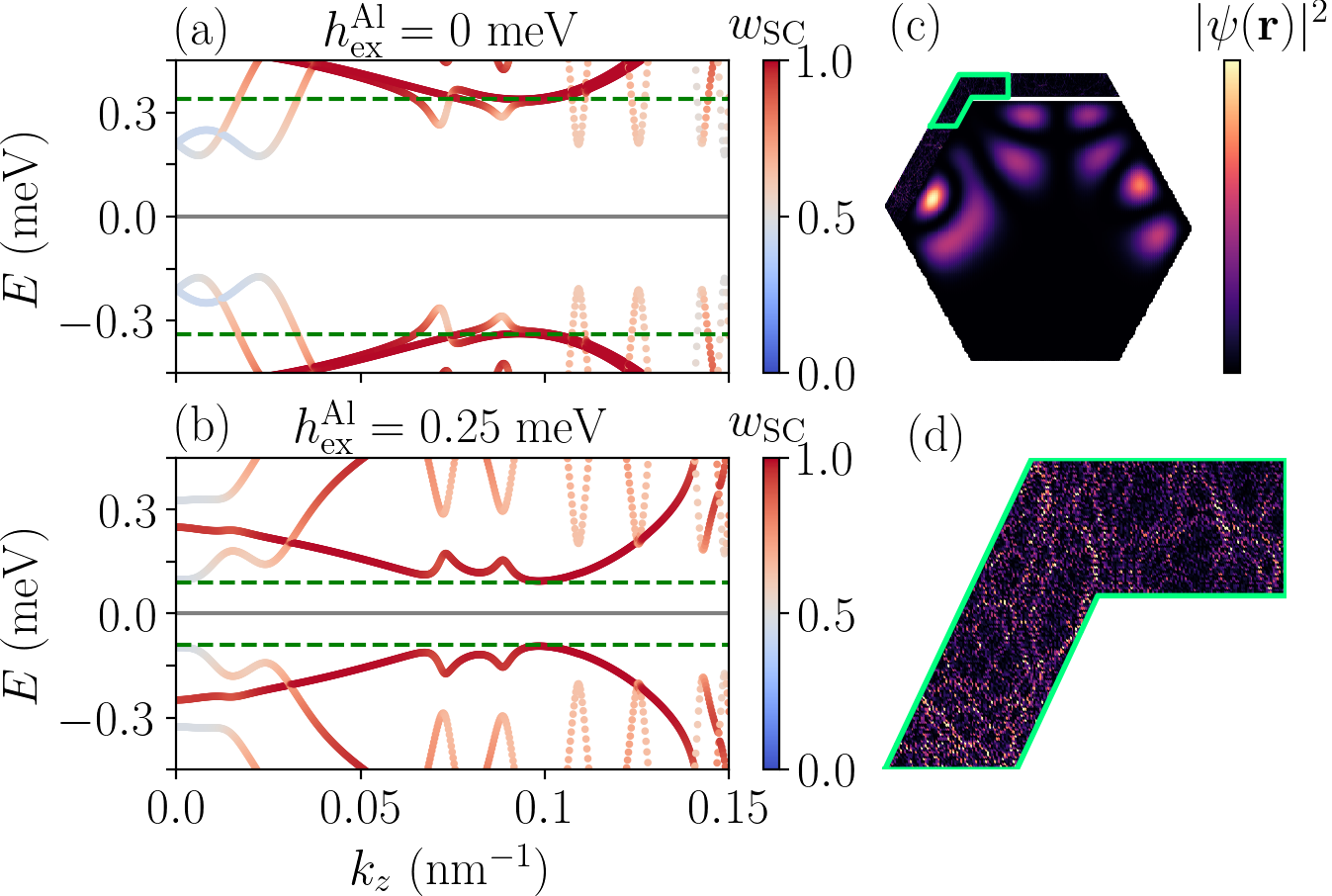}}
\caption{(a) and (b) BdG band diagrams for the InAs/Al hybrid nanowire in the absence and presence of the induced exchange coupling in Al. The gate voltages are fixed at $V_{\rm{BG}} = -3.4~$V, $V_{\rm{LG}} =V_{\rm{RG}}= 0~$V. We note that a finite $h^{\rm{Al}}_{\rm{ex}}$ lifts up the spin-orbit degeneracy at $k_z=0$ in the hybrid state and reduces the continuum gap of the superconducting states. (c) wavefunction profile of the hybrid state at $k_z=0$ and $E_{\rm{BdG}} \approx 0.2~$meV with $h^{\rm{Al}}_{\rm{ex}}=0$. (d) zoom-in of the wavefunction profile in the boxed region in Al (color scale adjusted). } 
\label{fig:band}
\end{figure}

\subsection{Exchange coupling in Al}
We first investigate the effect of induced exchange coupling inside the aluminum layer on the electronic properties of the InAs/Al hybrid system.
The origin of this exchange coupling is the magnetic proximity effect between the Al and EuS layers when they overlap with each other, as indicated in the schematic of Fig.~\ref{fig:schematic}.
To model this proximity effect, we assume that $h_{\text{ex}}(\bold{r}) = h^{\rm{Al}}_{\text{ex}} >0$ inside the Al layer. 
At this point we still neglect the magnetic proximity effect at the InAs/EuS interface; this will be discussed in the next section.

Figures~\ref{fig:band}(a) and~\ref{fig:band}(b) show the BdG band diagrams of the InAs/Al hybrid system in the absence ($h^{\rm{Al}}_{\text{ex}}$ = 0~meV) and presence ($h^{\rm{Al}}_{\text{ex}}$ = 0.25~meV) of the induced exchange coupling in Al, with the gate voltages being fixed at $V_{\rm{BG}} = -3.4~$V and $V_{\rm{LG}} = V_{\rm{RG}} = 0~$V.
The color of the band indicates the degree of wavefunction hybridization, which is defined as $w_{\text{SC}}=\sum_{\bold{r} \in \Omega_{\text{Al}}} | \psi(\bold{r})|^2 \leq 1$, with $\Omega_{\text{Al}}$ denoting the volume of the Al layer. 
A finite $h^{\rm{Al}}_{\text{ex}}$ has two effects on the band properties of the hybrid nanowire.

First, a finite $h^{\rm{Al}}_{\text{ex}}$ would induce an effective Zeeman spin slitting for the hybrid state.
As can be seen, the spin-orbit degeneracy at $k_z = 0$ and $E_{\rm{BdG}} \approx 0.2~$meV in Fig.~\ref{fig:band}(a) for the hybrid state ($w_{\text{SC}} \approx 0.5$) is now lifted by the finite induced exchange coupling in Al in Fig.~\ref{fig:band}(b). 
The amplitude of the effective Zeeman energy is approximately
\begin{align}
E^{(1)}_{Z}  \approx w_{\rm{SC}} \cdot h^{\rm{Al}}_{\text{ex}},
\label{eq:Ez_1}
\end{align}
which is proportional to the weight of the wavefunction in Al. 
Figures~\ref{fig:band}(c) and ~\ref{fig:band}(d) show the wavefunction profiles of the hybrid state in InAs and Al, respectively.  
Thereby, although InAs is not directly subject to the magnetic proximity effect from EuS in the physical scenario considered here, the hybrid state still gains a finite effective Zeeman spin splitting by distributing its wavefunction into the magnetized Al layer.

Second, the induced exchange coupling in Al would reduce the quasiparticle continuum gap.
By comparing those superconducting states ($w_{\text{SC}} \approx 1$) in Figs.~\ref{fig:band}(a) and ~\ref{fig:band}(b), we find that the excitation gap of the Al layer decreases from the bare value $\Delta_{\rm{qp}}=0.34~$meV to about $\Delta_{\rm{qp}} \approx 0.09~$meV [green dashed lines in Figs.~\ref{fig:band}(a) and ~\ref{fig:band}(b)].
Since Al is an $s$-wave BCS superconductivity, the quasiparticle continuum gap decreases with the induced exchange coupling in Al in a linear manner:
\begin{align}
\Delta_{\rm{qp}}(h^{\rm{Al}}_{\text{ex}}) = \Delta_0 - h^{\rm{Al}}_{\text{ex}}.
\label{eq:Delta_qp}
\end{align}
Thus we can estimate the strength of induced exchange coupling $h^{\rm{Al}}_{\text{ex}}$ from experimental data by considering the reduction of the quasiparticle continuum gap in Al.
On the other hand, for the hybrid state ($w_{\rm{SC}} \approx 0.5$), the $k_F$ excitation gap (inverse of the localization length of the Majorana modes) at $k_z \approx 0.025 \rm{nm}^{-1}$ in Figs.~\ref{fig:band}(a) and~\ref{fig:band}(b) changes very little with $h^{\rm{Al}}_{\text{ex}}$, possibly owing to the spin-orbit protection from InAs~\cite{Bommer2019SpinOrbit, Liu2019Conductance}.

When considering both of the abovementioned two effects on the InAs/Al hybrid nanowire, we conclude that an induced exchange coupling in Al alone \emph{cannot} drive the hybrid system into the topological phase.
Because by combining Eqs.~\eqref{eq:Ez_1} and~\eqref{eq:Delta_qp}, the induced effective Zeeman energy of the hybrid state  is always less than the induced superconducting gap, i.e., 
\begin{align}
E^{(1)}_{Z} < \Delta_{\rm{ind}} \approx w_{\rm{SC}}\Delta_0,
\label{eq:nogo}
\end{align}
as long as the quasiparticle continuum gap in Al remains finite $\Delta_{\rm{qp}}(h^{\rm{Al}}_{\text{ex}}) >0$.
This is in agreement with a fundamental no-go theorem for topology for BdG Hamiltonians \cite{Poyhonen2020}.

\begin{figure}
\centering
{\includegraphics[width = \linewidth]{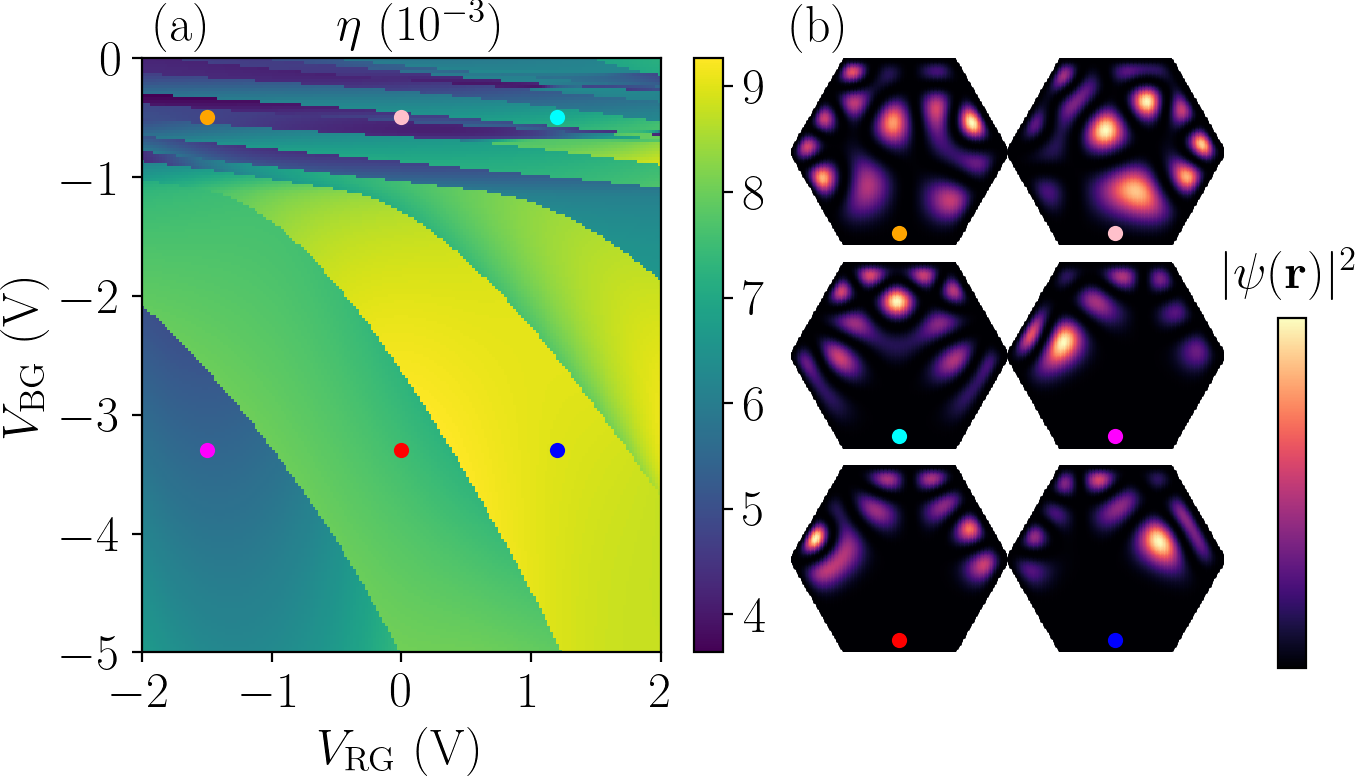}}
\caption{Magnetic proximity efficiency and wavefunction profiles in a bare InAs nanowire. (a) $\eta$ of the normal eigenstate closest to the Fermi surface as a function of the backgate and the rightgate voltages. (b) $|\psi(\bold{r})|^2$ of the normal eigenstates at specific gate voltages.}
\label{fig:magnetic}
\end{figure}

\subsection{Direct magnetic proximity effect}

We now focus on the direct magnetic proximity effect at the InAs/EuS interface and its dependence on gates, neglecting the superconducting shell completely. 
In particular for the quantum problem, we consider a bare InAs nanowire and the direct proximity effect is modeled phenomenologically as a local exchange coupling $h^{\rm{InAs}}_{\rm{ex}} \sigma_z$ within a distance $d=1.5~$nm from the two-facet boundaries where InAs and EuS contact with each other. 
Here, the distance $d$ is chosen to be about the penetration length of the wavefunction in a typical magnetic insulator~\cite{Sau2010NonAbelian}, such that the magnitude of $h^{\rm{InAs}}_{\rm{ex}}$ can be approximated as the strength of the ferromagnetic coupling inside EuS.
We have chosen for this phenomenological approach as the band structure of EuS may not be represented faithfully with an effective mass model as used for InAs and Al in our study.
The effect of the back-gate and two side-gates is included via the electrostatic potential profile $\phi(\bold{r})$, which is calculated based on the geometry shown in Fig.~\ref{fig:schematic}.
In order to quantify the magnetic proximity effect, we define the efficiency $\eta = [E_{n \su}(k_z=0) - E_{n \sd}(k_z=0)]/2h^{\rm{InAs}}_{\rm{ex}}$, which is the Zeeman energy splitting of the $n$-th spinful subband in the presence of a unit-strength $h^{\rm{InAs}}_{\rm{ex}}$. $E_{n \sigma}$ is the energy eigenstate of the discretized normal Hamiltonian $H_{\rm{N}}$ in Eq.~\eqref{eq:normal_ham}.

Figure~\ref{fig:magnetic}(a) shows the calculated $\eta$ of the normal subband mode closest to the Fermi surface as a function of the backgate and rightgate voltages (the leftgate dependence is weak due to the screening effect of Al).
The efficiency $\eta$ is a piecewise function of the gate voltages, with each piece corresponding to a particular subband mode. 
The $\eta$ difference between distinct subband modes can be stark and dominates the $\eta$ variations within a single subband mode.
Note that although the dependence of $\eta$ on the gate voltages is not monotonic, a general trend is that the subband mode at a more negative (positive) value of the backgate (rightgate) voltage would have a larger $\eta$, because their wavefunctions are more confined towards the InAs/EuS interface where the direct magnetic proximity effect is the strongest, as shown in Fig.~\ref{fig:magnetic}(b). 

The generalization from the bare InAs to the InAs/Al hybrid nanowire is straightforward.
Namely, the effective Zeeman splitting for the hybrid state due to the direct magnetic proximity effect can be approximated as
\begin{align}
E^{(2)}_{Z} \approx (1-w_{\rm{SC}} ) \cdot \eta \cdot h^{\rm{InAs}}_{\rm{ex}},
\label{eq:Ez_2}
\end{align}
where the prefactor $(1-w_{\rm{SC}} )$ accounts for the semiconductor-superconductor hybridization.
In the absence of other mechanisms of inducing Zeeman splitting, the minimal strength of the exchange coupling for realizing TSC would be about $h^{\rm{InAs}}_{\rm{ex}, \rm{min}} = \frac{w_{\rm{SC}} \Delta_0 }{(1-w_{\rm{SC}}) \eta}$ by requiring $E^{(2)}_{Z} = \Delta_{\rm{ind}}$. 
For a typical device with strong coupling at both InAs/Al and InAs/EuS interfaces, e.g., $w_{\rm{SC}} \approx 0.5$ and $\eta \approx 7 \times 10^{-3}$ [see Fig.~\ref{fig:magnetic}(a)], we have $h^{\rm{InAs}}_{\rm{ex, c} } \approx 50~$meV. 
Such a large strength of exchange coupling sets a demanding requirement for the proximity magnetic insulator candidates.

\begin{figure}
\centering
{\includegraphics[width = \linewidth]{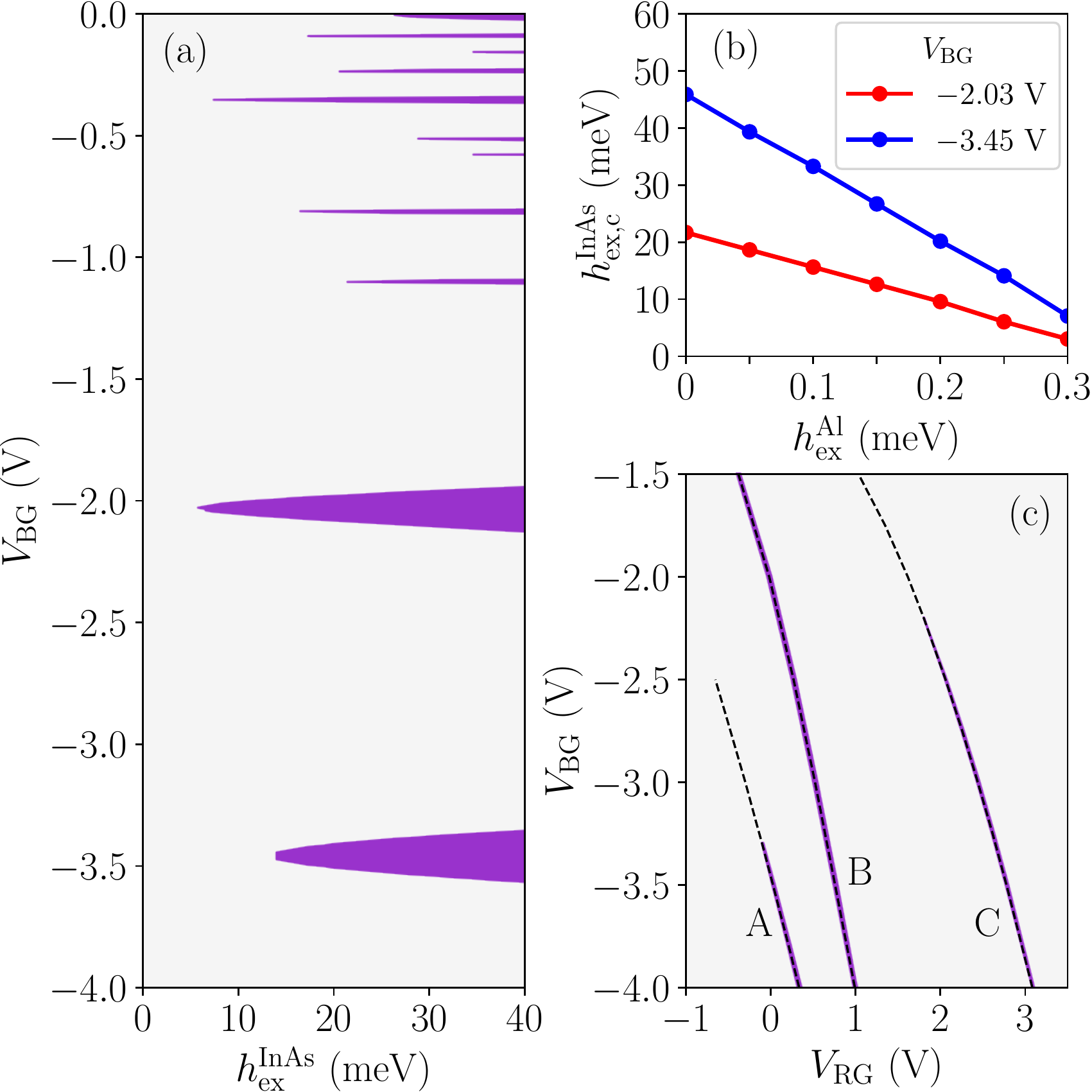}}
\caption{(a) Topological phase diagram in ($h^{\rm{InAs}}_{\rm{ex}}$, $V_{\rm{BG}}$) with $h^{\rm{Al}}_{\rm{ex}}=0.25~$meV, and $V_{\rm{LG}}=V_{\rm{RG}} = 0~$V. 
The area in purple represents the topological phase of the hybrid nanowire, while that in grey represents the trivial phase.
(b) Minimally required exchange coupling at the InAs/EuS interface for realizing TSC as a function of the strength of the induced exchange coupling in Al. 
The two lines correspond to the topological phases in (a) at $V_{\rm{BG}} = -2.03~$V and $-3.45~$V.
(c) Topological phase diagram in ($V_{\rm{RG}}$, $V_{\rm{BG}}$) with $h^{\rm{Al}}_{\rm{ex}}=0.25~$meV, $h^{\rm{InAs}}_{\rm{ex}}=15~$meV, and $V_{\rm{LG}} = 0~$V. 
}
\label{fig:phase_diagram} 
\end{figure}

\subsection{Topological phase diagram}

We now consider the scenario in which the InAs/Al hybrid nanowire is subject to the joint magnetic proximity effect from both Al/EuS and InAs/EuS interfaces, and study the topological phase diagrams as a function of gate voltages and exchange couplings. 
Namely, the induced exchange coupling is finite both in Al and at the boundaries of InAs, and thereby the total effective Zeeman spin splitting now is the combined contribution of two mechanisms:
\begin{align}
E^{\rm{tot}}_{Z}(h^{\rm{Al}}_{\rm{ex}}, h^{\rm{InAs}}_{\rm{ex}})= E^{(1)}_Z(h^{\rm{Al}}_{\rm{ex}}) + E^{(2)}_Z(h^{\rm{InAs}}_{\rm{ex}}),
\label{eq:Ez_tot}
\end{align}
where $E^{(1)}_Z$ and $E^{(2)}_Z$ are estimated in Eqs.~\eqref{eq:Ez_1} and~\eqref{eq:Ez_2}.
To determine the topological phase diagram of the hybrid nanowire, we keep track of the energy gap $E_{\rm{BdG}}(k_z=0)$.
For semiconductor-superconductor nanowires, the closing and reopening of $E_{\rm{BdG}}(k_z=0)$ signifies the topological quantum phase transition~\cite{Sau2010Generic, Lutchyn2010Majorana, Oreg2010Helical, Sau2010NonAbelian}.
Figure~\ref{fig:phase_diagram}(a) shows the topological phase diagram of the device in Fig.~\ref{fig:schematic} as a function of the backgate voltage $V_{\rm{BG}}$ and the exchange coupling $h^{\rm{InAs}}_{\rm{ex}}$ in InAs, with other parameters being fixed at $h^{\rm{Al}}_{\rm{ex}}=0.25~$meV, and $V_{\rm{LG}}=V_{\rm{RG}} = 0~$V.
The areas in purple represent the topological phase of the nanowire, while those in grey represent the trivial phase.
There are several observations on the result in Fig.~\ref{fig:phase_diagram}(a).
First, the pattern of the phase diagram resembles those of the hybrid nanowires for which the Zeeman energy is induced by an applied magnetic field but without including the orbital effect from the field.
Because in our model, the Zeeman energy is induced by the exchange couplings at zero magnetic field. 
Second, the TSC phases (lobes in purple) at $V_{\rm{RG}} < -1.5$~V are more robust, based on the fact that they have a smaller critical exchange coupling strength, and a larger width along $V_{\rm{BG}}$.
The robustness is the consequence of the joint effect of a larger direct magnetic proximity effect ($\eta > 7 \times 10^{-3}$ as shown in Fig.~\ref{fig:magnetic}) and a stronger InAs/Al hybridization ($w_{\rm{SC}} \approx 0.5$ as shown in Fig.~\ref{fig:band}) at more negative gate voltages.
Third, the minimal strength of the critical exchange field $h^{\rm{InAs}}_{\rm{ex, c}}$ for achieving topological phases is about 10~meV for the two lobes at $V_{\rm{RG}} \approx  -2$~V and $-3.5$~V.
Such a strength of $h^{\rm{InAs}}_{\rm{ex, c}}$ at the InAs/EuS interface is comparable to the estimated strength of exchange coupling at the interface of III-V compounds and magnetic insulators, which confirms the feasibility to realize TSC in semiconductor-superconductor-ferromagnetic hybrid nanowires with overlapping Al and EuS layers. 
This is one of the central result in the current work.

Figure~\ref{fig:phase_diagram}(b) shows the minimally required strength of $h^{\rm{InAs}}_{\rm{ex, c}}$ at the InAs/EuS interface as a function of $h^{\rm{Al}}_{\rm{ex}}$ in Al for two particular subband modes.
The minimal strength $h^{\rm{InAs}}_{\rm{ex, c}}$ decreases linearly with an increasing $h^{\rm{Al}}_{\rm{ex}}$, because an larger effective Zeeman energy $E^{(1)}_Z \propto h^{\rm{Al}}_{\rm{ex}}$ facilitates the realization of topological superconductivity in the hybrid nanowire.
In particular, the minimally required exchange coupling at the InAs/EuS interface is about $h^{\rm{InAs}}_{\rm{ex, c}} \sim $50 or 20~meV if no exchange coupling is induced in Al.
This value reduces significantly to $h^{\rm{InAs}}_{\rm{ex, c}} \lesssim $ 10 or 5~meV as $h^{\rm{Al}}_{\rm{ex}} \approx 0.28~$meV. 
Here for comparison between theory and experiment, the value of  $h^{\rm{Al}}_{\rm{ex}}$ is chosen such that the shrinking of the continuum gap is comparable to the observations in Ref.~\cite{Vaitiekenas2020Zerobias}, i.e., the gap in devices with overlapping Al and EuS layers is $\sim 0.04/0.23$ of the gap in non-overlapping ones. 
If we assume that the properties of a hybrid nanowire with non-overlapping Al and EuS layers are approximately captured by setting $h^{\rm{Al}}_{\rm{ex}}=0$ in our model, Fig.~\ref{fig:phase_diagram}(b) explains why zero-bias conductance peaks in the tunnel spectroscopy are only observed in overlapping devices in Ref.~\cite{Vaitiekenas2020Zerobias}.

Figure~\ref{fig:phase_diagram}(c) shows the topological phase diagram in the ($V_{\rm{RG}}$, $V_{\rm{BG}}$) plane, focusing on the three topological lobes at $V_{\rm{BG}} < -1.5~$V. 
Now the exchange couplings are fixed at $h^{\rm{InAs}}_{\rm{ex}}=15~$meV and $h^{\rm{Al}}_{\rm{ex}}=0.25~$meV, and gate voltages $V_{\rm{LG}}=0~\rm{V}$.
The topological phase shows up as a diagonal line, along which the Fermi energy of the relevant subband mode keeps close to zero.
Note that the hybrid state of the particular subband mode can remain topological all the way along the diagonal zero-Fermi-energy line (e.g., the continuous lobe-B), or it can transform between topologically trivial and nontrivial phases (e.g., lobes-A or -C).
It turns out that the topology along the zero-Fermi-energy line depends crucially on how the semiconductor-superconductor hybridization ($w_{\rm{SC}}$) and direct magnetic proximity efficiency ($\eta$) respond to the gate voltage variations.
For the hybrid state with zero Fermi energy, we can use a simplified criterion in the form
\begin{align}
& E^{\rm{tot}}_Z  - \Delta_{\rm{ind}}  \nn
=& E^{(2)}_Z - \left( \Delta_{\rm{ind}} - E^{(1)}_Z  \right) \nn
=& (1-w_{\rm{SC}} ) \cdot \eta \cdot h^{\rm{InAs}}_{\rm{ex}} - w_{\rm{SC}} (\Delta_0 - h^{\rm{Al}}_{\rm{ex}} ) > 0,
\label{eq:topol_Ef}
\end{align}
based on the definitions in Eqs.~\eqref{eq:Ez_1},~\eqref{eq:nogo},~\eqref{eq:Ez_2} and~\eqref{eq:Ez_tot}.
In Eq.~\eqref{eq:topol_Ef}, the relative strength of Zeeman energy due to the direct magnetic proximity effect $E^{(2)}_Z$ and the induced quasiparticle continuum gap $w_{\rm{SC}} (\Delta_0 - h^{\rm{Al}}_{\rm{ex}} ) $ depend on $w_{\rm{SC}}$ and $\eta$ explicitly.
%Put in another way, the effective Zeeman energy due to the direct magnetic proximity effect should be larger than the weighted superconducting quasiparticle continuum gap.

\begin{figure}
\centering
{\includegraphics[width = \linewidth]{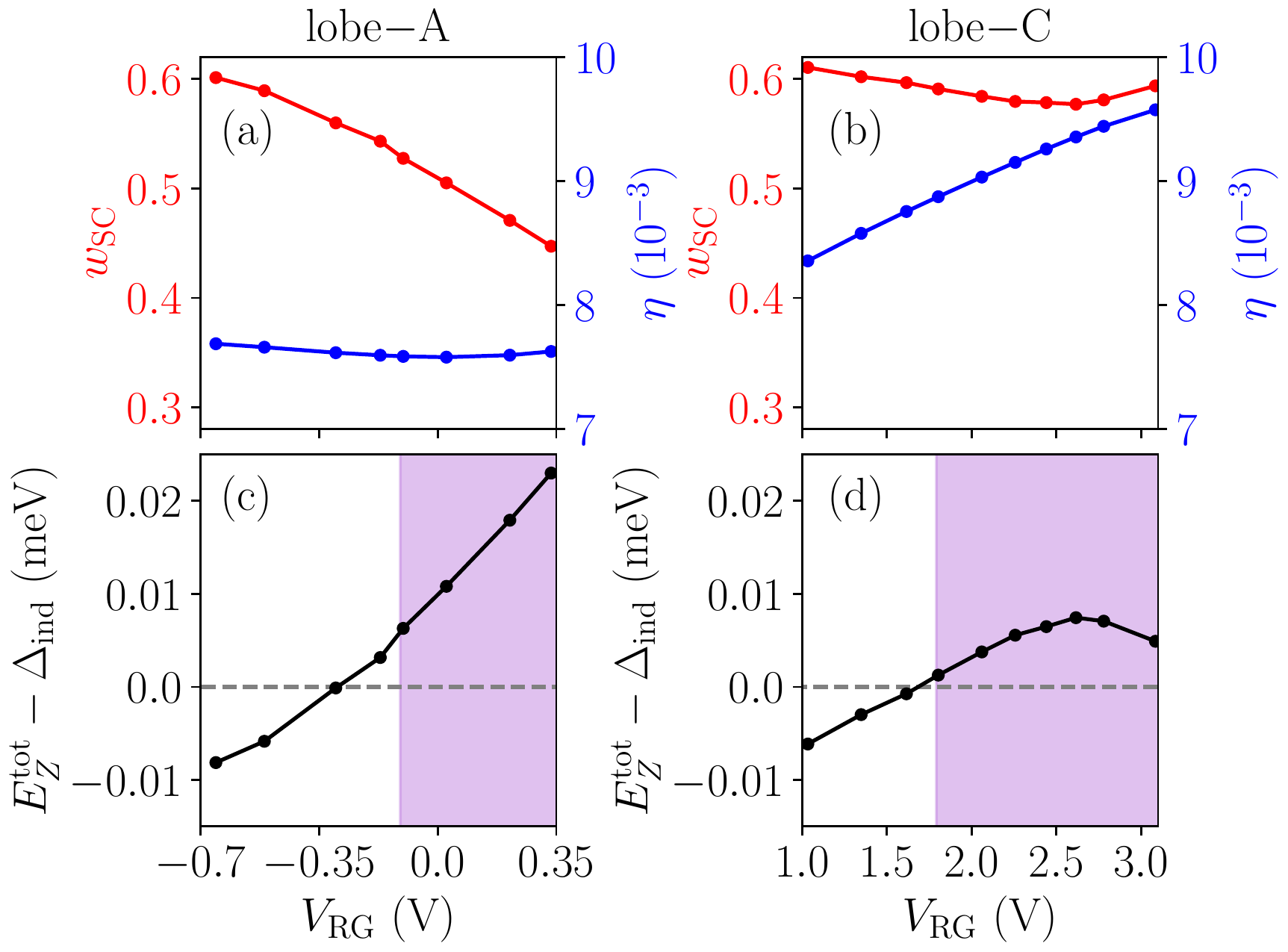}}
\caption{(a) $w_{\rm{SC}}$ and $\eta$ of lobe-A along the zero-Fermi-energy line [dashed lines in Fig.~\ref{fig:phase_diagram}(c)]. Here the variation of $w_{\rm{SC}}$ has a dominant effect over $\eta$ in determining the topological phase of the hybrid state. (c) Calculated $E^{\rm{tot}}_Z  - \Delta_{\rm{ind}}$ (black dots) of lobe-A. Ideally, the hybrid state at zero Fermi energy becomes topological when $E^{\rm{tot}}_Z  - \Delta_{\rm{ind}}$ is greater than zero. The purple shaded area represents the topological phase indicated in Fig.~\ref{fig:phase_diagram}(c). (b) and (d) Similar to (a) and (c) for lobe-C. For lobe-C, the change of $\eta$ is larger than $w_{\rm{SC}}$, and the hybrid state becomes topological when the direct magnetic proximity effect is prominent ($\eta > 9\times 10^{-3}$).   }
\label{fig:Ef_trace} 
\end{figure}

Figure~\ref{fig:Ef_trace} shows the $w_{\rm{SC}}$ and $\eta$ of the lobes-A and -C along the zero-Fermi-energy line, i.e., the dashed lines in Fig.~\ref{fig:phase_diagram}(c).
In Fig.~\ref{fig:Ef_trace}(a), the variation of $w_{\rm{SC}}$ dominates that of $\eta$, and the hybrid state is topological [see Fig.~\ref{fig:Ef_trace}(c)] when the hybridization is moderately small, i.e., $w_{\rm{SC}} \lesssim 0.5$.
As indicated by Eq.~\eqref{eq:topol_Ef}, a smaller degree of semiconductor-superconductor hybridization means a stronger $E^{(2)}_Z$ from the InAs side and a smaller induced continuum gap from Al, making it easier to satisfy the topological criterion.
In another scenario, as shown by Fig.~\ref{fig:Ef_trace}(b) for lobe-C, $\eta$ increases monotonically as the voltage of the right-gate becomes more positive, and has a dominant effect than $w_{\rm{SC}}$. 
The hybrid state becomes topological when $\eta$ is sufficiently large.
We thus see that depending on the details of a subband, a topological transition can be driven by two gates by both changing the induced superconducting gap or the directly induced Zeeman splitting.
This is in contrast to the usual topological phase transition driven by changing the chemical potential by a gate.

\section{Summary and discussions}

In this work, we studied the electronic properties of InAs/EuS/Al hybrid nanowires.
We analyzed the band bending at the InAs/EuS interface using ARPES data and found that this interface enhances electron accumulation compared to a bare InAs surface.
Using this input, we performed microscopic electrostatics and device simulations.
From these we concluded that it is feasible to achieve topological superconductivity in the device geometry shown in Fig.~\ref{fig:schematic}, within the realistic parameters:
the calculated minimal strength of $h^{\rm{InAs}}_{\rm{ex}}$ at the InAs/EuS interface is about $10~$meV, consistent with the induced exchange coupling between III-V semiconductors and magnetic insulators.
Our calculations also indicate that in experiments a topological phase is only achieved by the combination of both an induced Zeeman splitting in the superconducting Al shell by EuS, and an induced Zeeman splitting directly at the InAs/EuS interface. 
We also find in this hybrid device additional ways to control the topological phase by gates compared to the well-known control by changing the chemical potential:
Topology can be controlled using two gates either by changing the effective induced superconducting gap or by changing the overlap of the wave function with the InAs/EuS interface and thus the directly induced Zeeman splitting.
This gives new avenues to experimentally optimizing topological phases in a given device geometry.

While finishing this work we became aware of a similar study on InAs/EuS/Al nanodevices focusing on electrostatic effects \cite{Woods2020Electrostatic}.
That work concludes, opposite to our findings, that only the directly induced Zeeman splitting is necessary for a topological phase.
The reason for this discrepancy is that Ref.~\cite{Woods2020Electrostatic} only assumes electron accumulation due to the work function difference between Al and InAs, and not at the InAs/EuS interface, contrary to our experimental finding.
We note that there is concurrent work on the effects of electrostatics in these hybrid systems~\cite{LeviYeyati2020}. Also, there are concurrent efforts to go beyond the effective model as used in our work, and do a self-consistent treatment of proximity effect between EuS and Al when the shells overlap~\cite{Antipov2020}.

\begin{acknowledgements}
We are grateful to Aleksei Khindanov, Andrey E. Antipov, William S. Cole, Bernard van Heck for discussions at the initial stage of this project. 
We would like to thank Anton Akhmerov, Artem Pulkin, Haining Pan, and F. Setiawan for useful comments on the manuscript.
C.-X.L. thanks Zhenglu Li, Qisi Wang for helpful discussions.
S.S., Y.L. and P.K. would like to acknowledge J. Krieger and V. Strocov from the ADDRESS beamline at the Swiss Light Source, PSI, Switzerland.
This work was supported by a subsidy for top consortia for knowledge and innovation (TKl toeslag),
by the European Union's Horizon 2020 research and innovation programme FETOpen Grant No. 828948 (AndQC),
by Microsoft Quantum,
%by the European Research Council under the European Union's Horizon 2020 Research and innovation program (grant agreement no. 716655),
by the European Union's Horizon 2020 research and innovation programme under grant numbers 716655 (ERC Stg HEMs-DAM),
by the international training network ``INDEED'' (grant agreement no. 722176),
and by S$\tilde{a}$o Paulo Research Foundation, grants 2016/10167-8 and 2019/07082-9.
\end{acknowledgements}

\textit{Author contributions.}--C.-X.L. proposed the idea of microscopic device simulation and initiated the project, the scope of the project was later refined using contributions from all authors.
C.-X.L., K.V., and A.M. performed an extensive survey on the appropriate model of the hybrid nanowire. 
C.-X.L. conceived the model in this work, implemented the numerical methods, and performed the numerical calculations.
S.S., Y.L. and P.K. carried out the data analysis for band bending at the InAs/EuS interface.
M.W. supervised the project.
All authors discussed the results.
C.-X.L. and M. W. wrote the manuscript with input from all authors.

\bibliography{references.bib}

%\onecolumngrid
%\vspace{1cm}
%\begin{center}
%{\bf\large Supplemental Material for ``this paper"}
%\end{center}
%\vspace{0.5cm}

%\setcounter{secnumdepth}{3}
%\setcounter{equation}{0}
%\setcounter{figure}{0}
%\renewcommand{\theequation}{S-\arabic{equation}}
%\renewcommand{\thefigure}{S\arabic{figure}}
%\renewcommand\figurename{Supplementary Figure}
%\renewcommand\tablename{Supplementary Table}
%\newcommand\Scite[1]{[S\citealp{#1}]}
%\newcommand\Scit[1]{S\citealp{#1}}

%\makeatletter \renewcommand\@biblabel[1]{[S#1]} \makeatother
%%%%%%%%%%%%%%%%%%%%%%%%%%%%%%%%%%
% The supplementary text starts here
%%%%%%%%%%%%%%%%%%%%%%%%%%%%%%%%%%

%\section{xxxx}\label{sec:xxx}

\end{document}